\documentclass[prd,reprint,superscriptaddress,preprintnumbers,nofootinbib,amsmath,amssymb,aps]{revtex4-2}

\usepackage{graphicx}
\usepackage{dcolumn}
\usepackage{color}
\usepackage[colorlinks=true,citecolor=blue,urlcolor=blue]{hyperref}

\graphicspath{ {Figures/} }

\begin{document}

\preprint{UTTG-04-2021}

\title{Observational constraints on dark matter scattering with electrons}

\author{David Nguyen}
\author{Dimple Sarnaaik}
\affiliation{Department of Physics $\&$ Astronomy, University of Southern California, Los Angeles, CA, 90007, USA}

\author{Kimberly K.~Boddy}
\affiliation{Department of Physics, University of Texas at Austin, Austin, TX, 78712, USA}

\author{Ethan~O.~Nadler}
\affiliation{Kavli Institute for Particle Astrophysics and Cosmology and Department of Physics, Stanford University, Stanford, CA 94305, USA}
\affiliation{Carnegie Observatories, 813 Santa Barbara Street, Pasadena, CA 91101, USA}
\affiliation{Department of Physics $\&$ Astronomy, University of Southern California, Los Angeles, CA, 90007, USA}

\author{Vera Gluscevic}
\affiliation{Department of Physics $\&$ Astronomy, University of Southern California, Los Angeles, CA, 90007, USA}

\begin{abstract}
We present new observational constraints on the elastic scattering of dark matter with electrons for dark matter masses between 10 keV and 1 TeV. We consider scenarios in which the momentum-transfer cross section has a power-law dependence on the relative particle velocity, with a power-law index $n \in \{-4,-2,0,2,4,6\}$. We search for evidence of dark matter scattering through its suppression of structure formation. Measurements of the cosmic microwave background temperature, polarization, and lensing anisotropy from \textit{Planck} 2018 data and of the Milky Way satellite abundance measurements from the Dark Energy Survey and Pan-STARRS1 show no evidence of interactions. We use these data sets to obtain upper limits on the scattering cross section, comparing them with exclusion bounds from electronic recoil data in direct detection experiments. Our results provide the strongest bounds available for dark matter--electron scattering derived from the distribution of matter in the Universe, extending down to sub-MeV dark matter masses, where current direct detection experiments lose sensitivity.
\end{abstract}

\maketitle

\section{Introduction}

Cosmological observations are a powerful tool for studying the fundamental particle properties of dark matter (DM).
In the standard $\Lambda$CDM cosmology, DM is a cold, collisionless fluid.
However, if non-gravitational interactions between DM and ordinary matter exist, these interactions can have an observable effect on the distribution of matter throughout the Universe.

Elastic scattering between DM and baryons in the early Universe inhibits structure formation (with respect to $\Lambda$CDM), dampening the cosmic microwave background (CMB) anisotropies and suppressing the matter power spectrum on small scales~\cite{Boehm:2000gq,Chen:2002yh,Boehm:2004th}.
Previous studies have placed upper limits on the momentum-transfer cross section between DM and protons as a function of DM mass using measurements of CMB anisotropies from the \textit{Planck} satellite~\cite{Dvorkin:2013cea,Gluscevic:2017ywp,Boddy:2018kfv,Xu:2018efh,Slatyer:2018aqg,Boddy:2018wzy}.%
\footnote{There are also limits on DM scattering~\cite{Ali-Haimoud:2015pwa,Ali-Haimoud:2021lka} derived from the bounds on CMB spectral distortions~\cite{Fixsen:1996nj}.}
A variety of other observational probes of structure---including the Lyman-$\alpha$ forest~\cite{Viel:2013fqw,Irsic:2017ixq}, strong gravitational lensing~\cite{Hsueh:2019ynk,Gilman:2019nap}, stellar stream perturbations~\cite{Banik:2018pjp}, and Milky Way satellite galaxies~\cite{Kennedy:2013uta,Jethwa:2016gra,Nadler:2020prv,Newton:2020cog}---constrain the amount of suppression of the matter power spectrum at scales $\gtrsim 1~h~\mathrm{Mpc}^{-1}$.
Previous work has constrained DM--proton scattering using measurements of the Lyman-$\alpha$ forest power spectrum~\cite{Dvorkin:2013cea,Xu:2018efh} and, more recently, using the abundance of Milky Way satellite galaxies~\cite{Nadler:2019zrb,Nadler:2020prv,Maamari:2020aqz}.

These observational limits can be compared directly with the bounds from direct detection experiments searching for nuclear recoils, which cover complementary regions of parameter space for broad classes of DM models.
Such models can be described using low-energy effective field theory operators~\cite{Fan:2010gt,Fitzpatrick:2012ix,Anand:2013yka}; in a cosmological context, these operators produce momentum-transfer cross sections with a power-law dependence on the relative velocity between scattering DM particles and nucleons~\cite{Boddy:2018kfv}, permitting a straightforward comparison between constraints from cosmology and direct detection.
Direct detection experiments have achieved extraordinary sensitivity to the DM--nucleon cross section, primarily for DM masses above the GeV scale.
Cosmological observables probe much larger scattering cross sections, mostly outside the sensitivity range of direct detection experiments~\cite{Emken:2017qmp,Emken:2018run}, and DM masses $\gtrsim\mathrm{keV}$.

Cosmological studies of DM--baryon scattering have mainly focused on DM--proton scattering.
Observations can also provide bounds on DM--electron scattering, which are complementary to direct detection searches using electronic recoils~\cite{Essig:2012yx,Essig:2017kqs,Agnese:2018col,Agnes:2018oej,Crisler:2018gci,Aprile:2019xxb,Aguilar-Arevalo:2019wdi,Abramoff:2019dfb,Arnaud:2020svb,Barak:2020fql,Amaral:2020ryn}.
Electronic-recoil experiments have gained significant interest in recent years, because they can probe sub-GeV DM masses.
At present, the only cosmological constraints on DM--electron scattering are from CMB spectral distortions~\cite{Ali-Haimoud:2021lka}.

In this work, we focus on constraining DM--electron scattering using the latest measurements of the CMB temperature, polarization, and lensing anisotropies from the \textit{Planck} satellite~\cite{Aghanim:2019ame} and using the abundance of Milky Way satellites from the Dark Energy Survey (DES) and Pan-STARRS1~\cite{Drlica-Wagner:2019vah}.
We present constraints on the DM--electron momentum-transfer cross section for DM masses $\gtrsim 10$~keV, while electronic-recoil direct detection searches lose sensitivity below MeV mass scales.
Additionally, our limits extend to arbitrarily large cross sections,\footnote{Theoretical considerations place restrictions on the maximum DM cross section due to partial wave unitarity for point-like DM or finite-size considerations for composite DM. See Ref.~\cite{Digman:2019wdm} for a related discussion on DM--nucleus scattering.} while direct detection limits are subject to a detection ceiling~\cite{Lee:2015qva,Emken:2017erx,Emken:2019tni}.

In order to maintain the clear connection to direct detection experiments, we assume that DM scatters only with electrons and has no appreciable interaction with other Standard Model particles.
Such a scenario may arise in leptophilic models of DM~\cite{Bernabei:2007gr,Fox:2008kb,Baek:2008nz,Harnik:2008uu,Ibarra:2009bm,Dedes:2009bk,Cohen:2009fz,Cao:2009yy}, in which DM is not coupled to neutrinos~\cite{Harnik:2008uu,Dedes:2009bk,Chen:2018vkr}.
In more general frameworks, DM can scatter with various Standard Model particles.
Even in leptophilic models, there may be substantial DM--nucleon scattering induced at the loop level~\cite{Kopp:2009et,Bell:2014tta}.
Incorporating multiple scattering channels would strengthen cosmological constraints, but the relationship between the cross sections for different channels is model-dependent and left for future work.

During the completion of this manuscript, we learned of similar work in progress, presented in Ref.~\cite{Buen-Abad:2021}, which places constraints on DM--electron scattering for $n\in\{-4,-2,0\}$ using CMB and baryon acoustic oscillation (BAO) data, the abundance of Milky Way satellites, and the Lyman-$\alpha$ forest.
Where there is overlap, our results are in reasonable agreement, and we have verified that the inclusion of BAO data has little effect on our CMB constraints.
We note that Ref.~\cite{Buen-Abad:2021} includes an analysis of Milky Way satellites for $n=-2$ and $n=-4$.
Current methods~\cite{Nadler:2019zrb,Nadler:2020prv,Maamari:2020aqz} are not suitable for obtaining conservative limits for these cases, so we consider $n \geq 0$ only.
See Sec.~\ref{sec:analysis} for further discussion.

In Sec.~\ref{sec:theory}, we describe how the Boltzmann equations and cosmological observables are modified in the presence of DM--electron scattering.
In Sec.~\ref{sec:analysis}, we describe our procedure for constraining DM--electron scattering with \textit{Planck} data and with Milky Way satellite abundance data, and we present our results.
In Sec.~\ref{sec:compare}, we compare our bounds with limits from direct detection experiments, for selected models.
We conclude in Sec.~\ref{sec:conclusions}.

\section{Dark matter scattering}
\label{sec:theory}

Elastic scattering between DM and ordinary matter in the early Universe transfers energy and momentum between the DM and baryon fluids, suppressing the formation of structure at progressively smaller scales.
This suppression dampens the small-scale CMB power spectra and may, depending on the scattering model, create a sharp cutoff in the matter power spectrum (with respect to $\Lambda$CDM) at small scales~\cite{Chen:2002yh}.

When working with the cosmological Boltzmann equations, electrons are treated as a component of the nonrelativistic baryon fluid due to their tight coupling to baryonic particles.
The treatment of DM scattering with electrons rather than protons or helium is a matter of DM scattering with a different component of the baryon fluid, with constituent particles of a different mass.

\subsection{Models}
\label{sec:theory-models}

The relevant scattering quantity entering the Boltzmann equations in Sec.~\ref{sec:theory-boltzmann} is the momentum-transfer scattering cross section, obtained by weighting the differential cross section by the fractional longitudinal momentum transferred in the scattering process:
\begin{equation}
  \sigma_\textrm{MT} \equiv \int d\Omega \frac{d\sigma}{d\Omega} (1-\cos\theta) \, ,
  \label{eq:sigmaMT}
\end{equation}
where $\theta$ is the scattering angle.
We parameterize this cross section as
\begin{equation}
  \sigma_\textrm{MT} = \sigma_0 v^n \, ,
  \label{eq:sigmaMT-param}
\end{equation}
where $\sigma_0$ is a constant coefficient and $v$ is the relative velocity between the incoming scattering particles with a power-law index $n$.

This parameterization of the velocity dependence encompasses a wide class of DM models.
In an effort to be agnostic towards the underlying UV theory of DM, we may consider effective field theories that allow DM and electrons to interact through higher-dimensional operators~\cite{Kopp:2009et,Chang:2014tea,Rawat:2017fak,Bishara:2018vix}.
Since we are concerned with DM interactions in the nonrelativistic regime, we can adapt the nonrelativistic operators formalism for DM--nucleon scattering~\cite{Fan:2010gt,Fitzpatrick:2012ix,Anand:2013yka} to the case of DM--electron scattering~\cite{Catena:2019gfa}; these nonrelativistic operators map onto linear combinations of the relativistic operators.
Reference~\cite{Boddy:2018kfv} showed how these nonrelativistic operators are cast into the form of Eq.~\eqref{eq:sigmaMT-param} for use in a cosmological setting, and the possible velocity dependencies are $n\in\{0,2,4,6\}$, assuming no additional velocity- or momentum-dependence is introduced through the Wilson coupling coefficients.

Negative values of $n$ arise when DM interacts with electrons through a very light mediator, with a mass much smaller than the typical amount of momentum transferred via scattering.
For example, DM with an electric dipole moment scatters with $n=-2$~\cite{Sigurdson:2004zp}.
The case of $n=-4$ is relevant for millicharged DM, in which DM possesses a small electric charge that permits Coulomb interactions (e.g., see Ref.~\cite{Dolgov:2013una,Kovetz:2018zan,dePutter:2018xte}).
We note that DM interacting with electrons through an electromagnetic channel would also permit interactions with other charged particles, such as protons and helium nuclei.
In this work, we purposefully limit our scope to DM--electron scattering only in order to make fair comparisons with electronic-recoil direct detection experiments.
An analysis of any particular model with multiple scattering channels would strengthen the results we present in Sec.~\ref{sec:analysis}.

\subsection{Boltzmann equations}
\label{sec:theory-boltzmann}

In the presence of interactions between the DM (denoted as $\chi$) and baryon (denoted as $b$) fluids, a collision term in the Boltzmann equations couples the motion and temperature of the two fluids.
The standard Boltzmann equations of $\Lambda$CDM~\cite{Ma:1995ey} are modified to be~\cite{Chen:2002yh}
\begin{align}
  \dot{\delta}_b &= -\theta_b - \frac{\dot{h}}{2}, \
  \dot{\delta}_\chi = -\theta_\chi - \frac{\dot{h}}{2} \nonumber\\
  \dot{\theta}_b &= -\frac{\dot{a}}{a}\theta_b   + c_b^2 k^2 \delta_b + R_\gamma (\theta_\gamma - \theta_b)+ \frac{\rho_\chi}{\rho_b} R_\chi (\theta_\chi - \theta_b) \nonumber\\
  \dot{\theta}_\chi &= -\frac{\dot{a}}{a}\theta_\chi + c_\chi^2 k^2 \delta_\chi + R_\chi (\theta_b - \theta_\chi) \, ,
\label{eq:boltzmann}
\end{align}
where $\delta_{\chi,b}$ and $\theta_{\chi,b}$ are the density fluctuations and velocity divergences, respectively, of the fluids in Fourier space; $c_{\chi,b}$ are the speeds of sound in the fluids; and $\rho_{\chi,b}$ are their energy densities.
The overdot represents a derivative with respect to conformal time, $k$ is the wave number of a given Fourier mode, $a$ is the scale factor, and $h$ is the trace of the scalar metric perturbation.
The temperatures of the fluids evolve as%
\footnote{In this work, we do not incorporate the backreaction of DM scattering on the evolution of the baryon temperature; however, its effect on the CMB power spectra is subdominant and should have little impact on our analysis results, as investigated in Ref.~\cite{Boddy:2018wzy}.}
\begin{align}
  \dot{T}_b + 2\frac{\dot{a}}{a} T_b &= 2\frac{\mu_b}{m_e} R_\gamma (T_\gamma - T_b) + 2\frac{\mu_b}{m_\chi} R^\prime_\chi (T_\chi - T_b) \nonumber\\
  \dot{T}_\chi + 2\frac{\dot{a}}{a} T_\chi &= 2R^\prime_\chi (T_b - T_\chi) \, ,
  \label{eq:temperature}
\end{align}
where $m_e$ is the mass of the electron, $m_\chi$ is the mass of the DM particle, $\mu_b$ is the mean molecular weight of the baryons, and $T_\gamma$ is the photon temperature.

The terms proportional to $R_\gamma$ and $R_\chi$ in Eq.~\eqref{eq:boltzmann} describe the transfer of momentum between interacting fluids, acting as a drag force between the fluids.
The momentum-transfer rate coefficient $R_\gamma$ arises from Compton scattering between photons and electrons.
The rate coefficient for DM--electron scattering is
\begin{equation}
  R_\chi = a \rho_e \frac{\mathcal{N}_n \sigma_0}{m_\chi + m_e}
  \left(\frac{T_\chi}{m_\chi} + \frac{T_b}{m_e} \right)^{(n+1)/2} \, ,
  \label{eq:rate}
\end{equation}
where $\mathcal{N}_n \equiv 2^{(5+n)/2} \Gamma(3+n/2)/(3\sqrt{\pi})$, $\rho_e = (1-Y_\mathrm{He}) \rho_b x_e m_e/m_p$ is the electron density, $Y_\mathrm{He}$ is the helium mass fraction, $m_p$ is the proton mass, and $x_e$ is the ionization fraction.
This expression has a similar form seen in previous CMB literature on DM--proton scattering~\cite{Dvorkin:2013cea,Gluscevic:2017ywp,Boddy:2018kfv,Xu:2018efh,Slatyer:2018aqg,Boddy:2018wzy}, except the mass and density of protons are substituted for the mass and density of electrons.
The heat-transfer rate coefficient in Eq.~\eqref{eq:temperature} is $R^\prime_\chi = R_\chi m_\chi / (m_\chi + m_e)$.

In deriving Eq.~\eqref{eq:rate}, we assume DM particles possess a Maxwell-Boltzmann distribution function.
Following the current standard of cosmological analyses, we neglect possible deviations in the distribution function induced by DM scattering; as a result, our analysis may overestimate the constraining power on DM scattering by a factor of a few for low-mass DM and for DM cross sections with a steep velocity dependence~\cite{Ali-Haimoud:2018dvo}, but a detailed analysis is required.
It is also possible that DM is produced with a nontrivial distribution function, as is the case for freeze-in DM, which exhibits $n=-4$ scattering~\cite{Dvorkin:2019zdi,Dvorkin:2020xga}.
We do not consider such scenarios in this work.

The evolution equations in Eqs.~\eqref{eq:boltzmann} and \eqref{eq:temperature} are valid at linear order, assuming the relative bulk velocity between the DM and baryon fluids is small compared to the thermal relative velocity between scattering particles $v_\mathrm{th} = (T_\chi/m_\chi + T_b/m_e)^{1/2}$, which appears in the rate coefficient in Eq.~\eqref{eq:rate}.
For models with $n \geq 0$, the momentum-transfer rate is large at early times, which efficiently couples the motion of the DM and baryon fluids, rendering the relative bulk velocity small prior to recombination.
The rate coefficient $R_\chi$ given in Eq.~\eqref{eq:rate} is appropriate for these cases.

For $n=-2$ and $n=-4$, the DM scattering rate is feeble in the early Universe, and the relative bulk velocity can exceed the thermal velocity at times relevant for the CMB.
As a result, the Boltzmann equations become nonlinear~\cite{Dvorkin:2013cea,Boddy:2018wzy}.
In order to account for this nonlinearity, we follow Refs.~\cite{Dvorkin:2013cea,Xu:2018efh,Slatyer:2018aqg} to modify Eq.~\eqref{eq:rate} with the substitution
\begin{equation}
  \left(\frac{T_\chi}{m_\chi} + \frac{T_b}{m_e}\right) \to
  \left(\frac{T_\chi}{m_\chi} + \frac{T_b}{m_e} + \frac{V_\textrm{RMS}^2}{3}\right) \, ,
  \label{eq:modification}
\end{equation}
where we approximate the bulk velocity to be its root-mean-square $V_\textrm{RMS}(z)$ under $\Lambda$CDM: $V_\textrm{RMS} \sim 30~\mathrm{km/s}$ at $z\gtrsim 10^3$ prior to recombination and evolves as $(1+z)^2$ at later times~\cite{Tseliakhovich:2010bj}.
We note that using $V_\textrm{RMS}$ from a $\Lambda$CDM cosmology is a good approximation to its value in a cosmology where 100\% of DM is interacting and the interaction strength is no larger than its current CMB bounds; however, this ceases to be the case if only a fraction of DM interacts with baryons~\cite{Boddy:2018wzy}.
We only consider the former case.

\section{Analysis and results}
\label{sec:analysis}

\begin{figure*}[t]
  \includegraphics[width=0.48\linewidth]{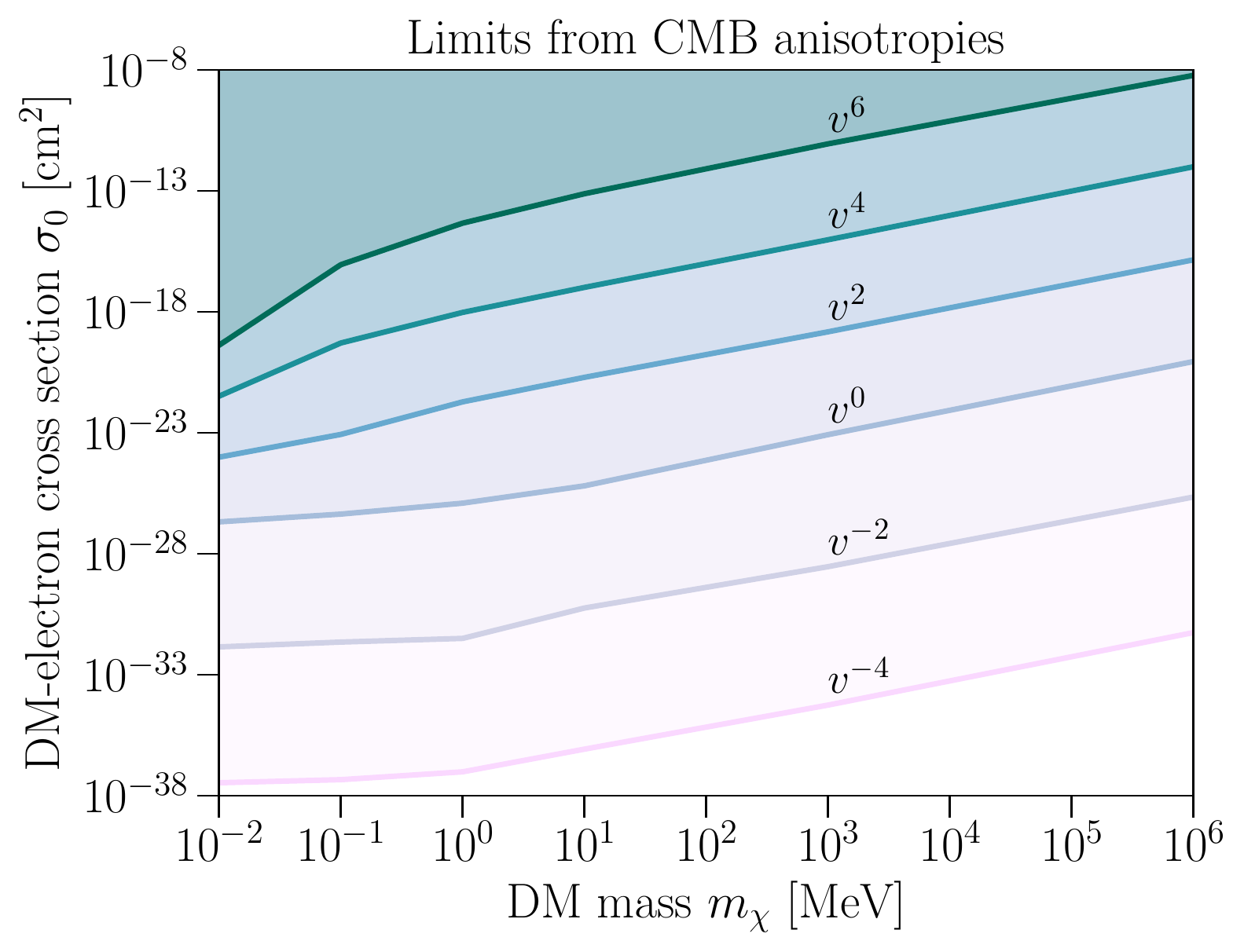}
  \includegraphics[width=0.48\linewidth]{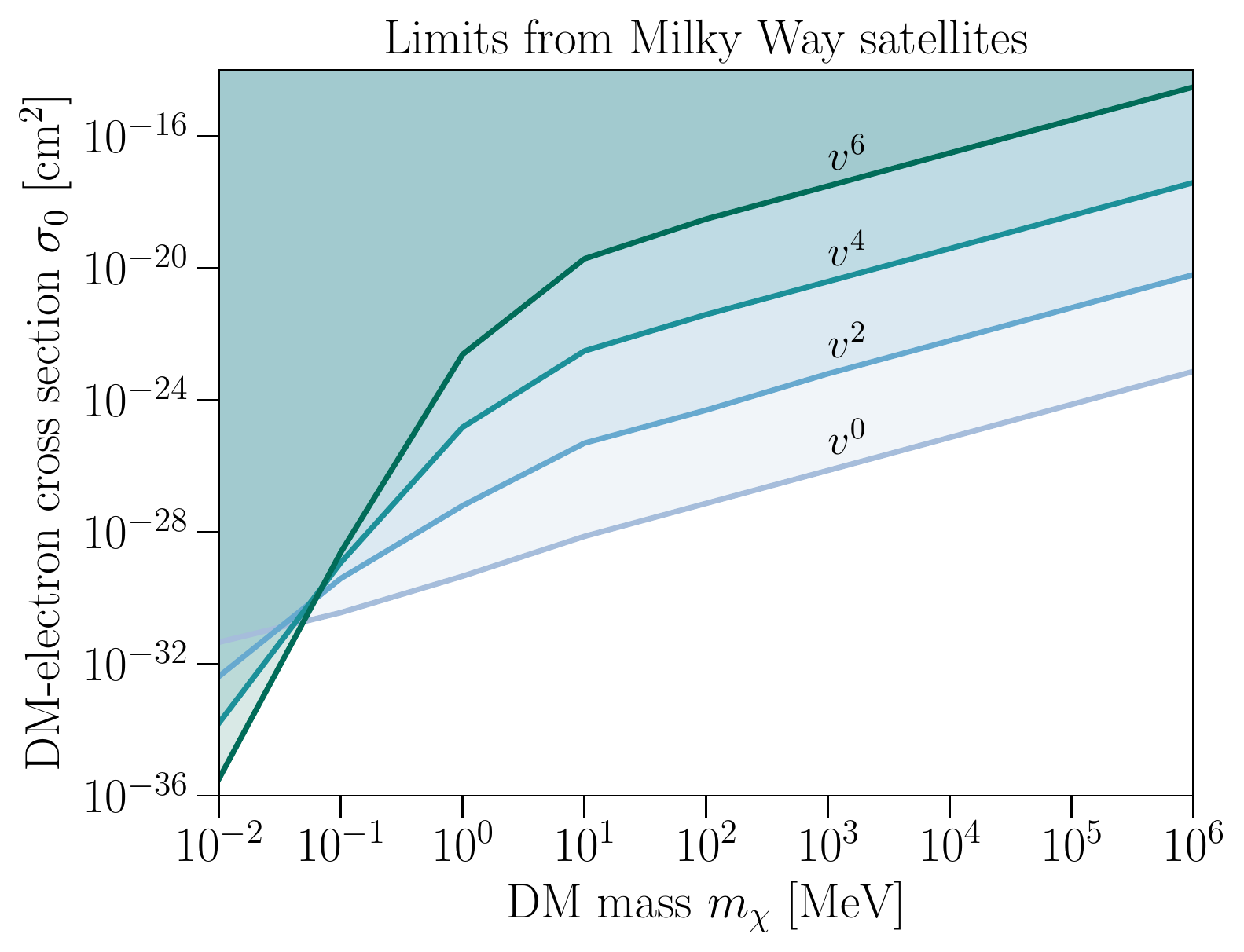}
  \caption{The 95\% C.L.\ upper limits on $\sigma_0$, the coefficient of the momentum-transfer cross section for DM--electron scattering, as a function of DM mass.
    The cross section scales as $v^n$, as indicated in the plots, where $v$ is the relative velocity of scattering particles.
    The shaded region above each line is excluded by (\textbf{left}) \textit{Planck} 2018 CMB temperature, polarization, and lensing power spectra at the 95\% C.L.\ and (\textbf{right}) the Milky Way satellite abundance from DES and Pan-STARRS1.}
  \label{fig:limits_all}
\end{figure*}

In this section, we describe our analysis methods for constraining DM--electron scattering.
For both analyses, we place upper bounds on the coefficient $\sigma_0$ of the DM--electron momentum-transfer cross section as a function of the DM mass in the range $10~\mathrm{keV} \leq m_\chi \leq 1~\mathrm{TeV}$.

We choose the lower end of the DM mass range to be $10~\mathrm{keV}$, because the validity of our assumptions for thermalized, cold DM breaks down at smaller masses for $n \geq 0$.%
\footnote{For $n<0$, the DM temperature is lower than the photon-baryon temperature~\cite{Boddy:2018wzy}, possibly allowing our analysis to extend to lower masses. Therefore, we also provide results for a $1~\mathrm{keV}$ DM mass in the tables below.}
At large DM masses $m_\chi \gg m_e$, the rate coefficient in Eq.~\eqref{eq:rate} becomes a function of $\sigma_0 / m_\chi$, allowing extrapolation of the form $\sigma_0 \propto m_\chi$ for all of our results to larger DM masses.
For practicality, we limit our analysis to a maximum DM mass of 1~TeV.

We use a modified version\footnote{\url{https://github.com/kboddy/class_public/tree/dmeff}} of the Cosmic Linear Anisotropy Solving System (\texttt{CLASS}) code\footnote{\url{https://github.com/lesgourg/class_public}} to compute the CMB power spectra and linear matter power spectrum within a cosmology that features DM--electron scattering~\cite{Gluscevic:2017ywp,Boddy:2018kfv,Boddy:2018wzy}.

\subsection{CMB}
\label{sec:analysis-cmb}

We perform a Markov chain Monte Carlo likelihood analysis of the \textit{Planck} 2018 CMB temperature, polarization, and lensing power spectra~\cite{Aghanim:2019ame}, using the \texttt{planck\_2018\_highl\_plik.TTTEEE\_lite} and \texttt{planck\_2018\_lensing.clik} likelihoods, in order to place upper bounds on the DM--electron momentum-transfer cross section.
We consider a set of seven cosmological parameters ${\theta}=\left\{ n_s, \, \tau_{\mathrm{reio}}, \, \log \left( 10^{10}A_s \right), \, \Omega_b h^2, \, \Omega_c h^2, \, 100 \theta_s,  \, \sigma_{\mathrm{0}} \right\}$, representing the standard six $\Lambda$CDM parameters and the momentum-transfer cross section coefficient $\sigma_0$ for DM--electron elastic scattering.
Following \textit{Planck}~\cite{Aghanim:2018eyx}, we assume three standard neutrino species, represented by two massless states and one $0.06~\mathrm{eV}$ massive state.

When computing the CMB power spectra using \texttt{CLASS}, we do not incorporate any nonlinear effects.
Currently available tools for calculating the nonlinear growth of perturbations, such as \texttt{Halofit}, are not reliable in context of cosmologies featuring DM scattering with baryons~\cite{Li:2018zdm}.
Furthermore, nonlinear growth amplifies perturbations on scales smaller than those directly relevant for \textit{Planck}.
Thus, we assume that linear cosmology describes our data sufficiently well and leave studies of nonlinearities in interacting cosmologies for future work.

To sample the posterior probability distribution of $\theta$, we use the publicly-available \texttt{Cobaya}\footnote{\url{https://cobaya.readthedocs.io/en/latest/index.html}} framework.
We employ broad flat priors on all parameters.
In each likelihood analysis, we fix the DM scattering model by the choice of the power-law index $n\in\{-4, -2, 0, 2, 4, 6\}$.
To speed up convergence in the sampling process, we fix the DM mass $m_\chi$ for each run and compute chains of parameter samples for various benchmark masses.
The resulting  95\% confidence level (C.L.) upper limits on $\sigma_0$ for each sampled mass are shown in the left panel of Fig.~\ref{fig:limits_all} and listed in Table~\ref{tab:bounds-cmb}.

\subsection{Milky Way satellites}
\label{sec:analysis-mw}

Reference~\cite{Nadler:2019zrb} demonstrated that velocity-independent DM--proton scattering in the early Universe produces a suppression of the linear matter power spectrum and the corresponding subhalo mass function in a very similar manner to cosmologies with WDM.
This similarity enables a correspondence between the DM--proton momentum-transfer cross section and WDM mass, which was used to place constraints on the scattering scenario via a WDM analysis~\cite{Nadler:2019zrb,Nadler:2020prv}.

For models in which $n \geq 0$, the rate of momentum transfer is larger at higher redshifts and, for most purposes, negligible after recombination, provided that the interaction strength is below current cosmological bounds~\cite{Dvorkin:2013cea,Boddy:2018kfv,Maamari:2020aqz}.
For this reason, it is possible to capture the effects of such interaction models on the population of satellite galaxies by considering only their effects on the transfer function (i.e., the ratio of the linear matter power spectrum $P(k)$ in a modified cosmology to that in a $\Lambda$CDM cosmology).

\begin{table}[t]
  \centering
  \begin{tabular}{|r!{}l|c|c|c|c|c|c|}
    \hline
    \multicolumn{2}{|c|}{\textbf{DM}} & \multicolumn{6}{c|}{$n$} \\
    \cline{3-8}
    \multicolumn{2}{|c|}{\textbf{mass}} & $-4$ & $-2$ & $0$ & $2$ & $4$ & $6$ \\
    \hline
    1 & keV & 4.0e-38 & 1.7e-32 & 1.1e-27 & 1.1e-25 & 2.7e-24 & 5.1e-23 \\
    \hline
    10 & keV & 3.4e-38 & 1.4e-32 & 2.1e-27 & 9.9e-25 & 3.3e-22 & 4.2e-20 \\
    \hline
    100 & keV & 4.6e-38 & 2.3e-32 & 4.4e-27 & 8.6e-24 & 5.2e-20 & 8.9e-17 \\
    \hline
    1 & MeV & 9.7e-38 & 3.2e-32 & 1.2e-26 & 1.9e-22 & 9.5e-19 & 4.7e-15 \\
    \hline
    10 & MeV & 8.4e-37 & 5.7e-31 & 6.5e-26 & 2.0e-21 & 1.0e-17 & 7.7e-14 \\
    \hline
    1 & GeV & 5.6e-35 & 2.9e-29 & 8.4e-24 & 1.5e-19 & 9.6e-16 & 8.8e-12 \\
    \hline
    1 & TeV & 5.6e-32 & 2.2e-26 & 8.9e-21 & 1.4e-16 & 1.0e-12 & 6.0e-09 \\
    \hline
  \end{tabular}
  \caption{The 95\% C.L.\ upper limits on $\sigma_0$, the coefficient of the momentum-transfer cross section for DM--electron scattering, in units of $\mathrm{cm}^2$ from the CMB analysis of Sec.~\ref{sec:analysis-cmb} and shown in the left panel of Fig.~\ref{fig:limits_all}.}
  \label{tab:bounds-cmb}
\end{table}

\begin{table}[t]
  \centering
  \begin{tabular}{|r!{}l|c|c|c|c|}
    \hline
    \multicolumn{2}{|c|}{\textbf{DM}} & \multicolumn{4}{c|}{$n$} \\
    \cline{3-6}
    \multicolumn{2}{|c|}{\textbf{mass}} & 0 & 2 & 4 & 6 \\
    \hline
    1 & keV & 5.7e-33 & 4.2e-36 & 1.6e-39 & 3.3e-43 \\
    \hline
    10 & keV & 4.5e-32 & 4.2e-33 & 1.6e-34 & 3.3e-36 \\
    \hline
    100 & keV & 3.6e-31 & 3.9e-30 & 1.2e-29 & 2.4e-29 \\
    \hline
    1 & MeV & 4.5e-30 & 6.2e-28 & 1.5e-25 & 2.4e-23 \\
    \hline
    10 & MeV & 7.3e-29 & 4.9e-26 & 3.0e-23 & 1.9e-20 \\
    \hline
    100 & MeV & 7.3e-28 & 4.9e-25 & 3.9e-22 & 3.0e-19 \\
    \hline
    1 & GeV & 7.3e-27 & 6.2e-24 & 3.9e-21 & 3.0e-18 \\
    \hline
    10 & GeV & 7.3e-26 & 6.2e-23 & 3.9e-20 & 3.0e-17 \\
    \hline
    100 & GeV & 7.3e-25 & 6.2e-22 & 3.9e-19 & 3.0e-16 \\
    \hline
    1 & TeV & 7.3e-24 & 6.2e-21 & 3.9e-18 & 3.0e-15 \\
    \hline
  \end{tabular}
  \caption{Conservative upper limits on $\sigma_0$, the coefficient of the momentum-transfer cross section for DM--electron scattering, in units of $\mathrm{cm}^2$ from the Milky Way satellite analysis of Sec.~\ref{sec:analysis-mw} and shown in the right panel of Fig.~\ref{fig:limits_all}.}
  \label{tab:bounds-mw}
\end{table}

In addition to small-scale suppression, the efficient coupling between the DM and baryon fluids generates dark acoustic oscillations in the linear matter power spectrum.
For $n=0$, the dark acoustic oscillations are negligible at the scattering limit found in Ref.~\cite{Nadler:2019zrb}, and the matter power spectrum features a WDM-like cutoff.
For velocity-dependent scattering with $n>0$, the dark acoustic oscillations are substantial below the cutoff scale, and the recovery of power at very small scales invalidates the direct correspondence with WDM.

To address the dark acoustic oscillations in models with $n>0$, Ref.~\cite{Maamari:2020aqz} developed a general and very conservative numerical procedure for mapping WDM constraints to limits on DM--proton scattering by comparing the respective transfer functions.
Namely, for a given $m_{\chi}$ and $n$, the strength of DM--proton scattering is considered strictly ``ruled out'' if the suppression of the transfer function is more severe than that of thermal relic WDM, at the current lower limit on its mass, up to a very large $k$.
This approach yielded the strongest observational limits on models for velocity-dependent scattering with protons, and we adopt it here for the case of electron scattering.
For this purpose, we use the lower limits on the mass of WDM of $6.5~\mathrm{keV}$ (at $95\%$ confidence) reported by Ref.~\cite{Nadler:2020prv}, which relied on the measurements of the abundance of Milky Way satellites over nearly the full sky, including the population of satellites accreted with the Large Magellanic Cloud, detected in DES and Pan-STARRS1 data~\cite{Drlica-Wagner:2019vah}.

\begin{figure}[t]
  \includegraphics[width=0.98\linewidth]{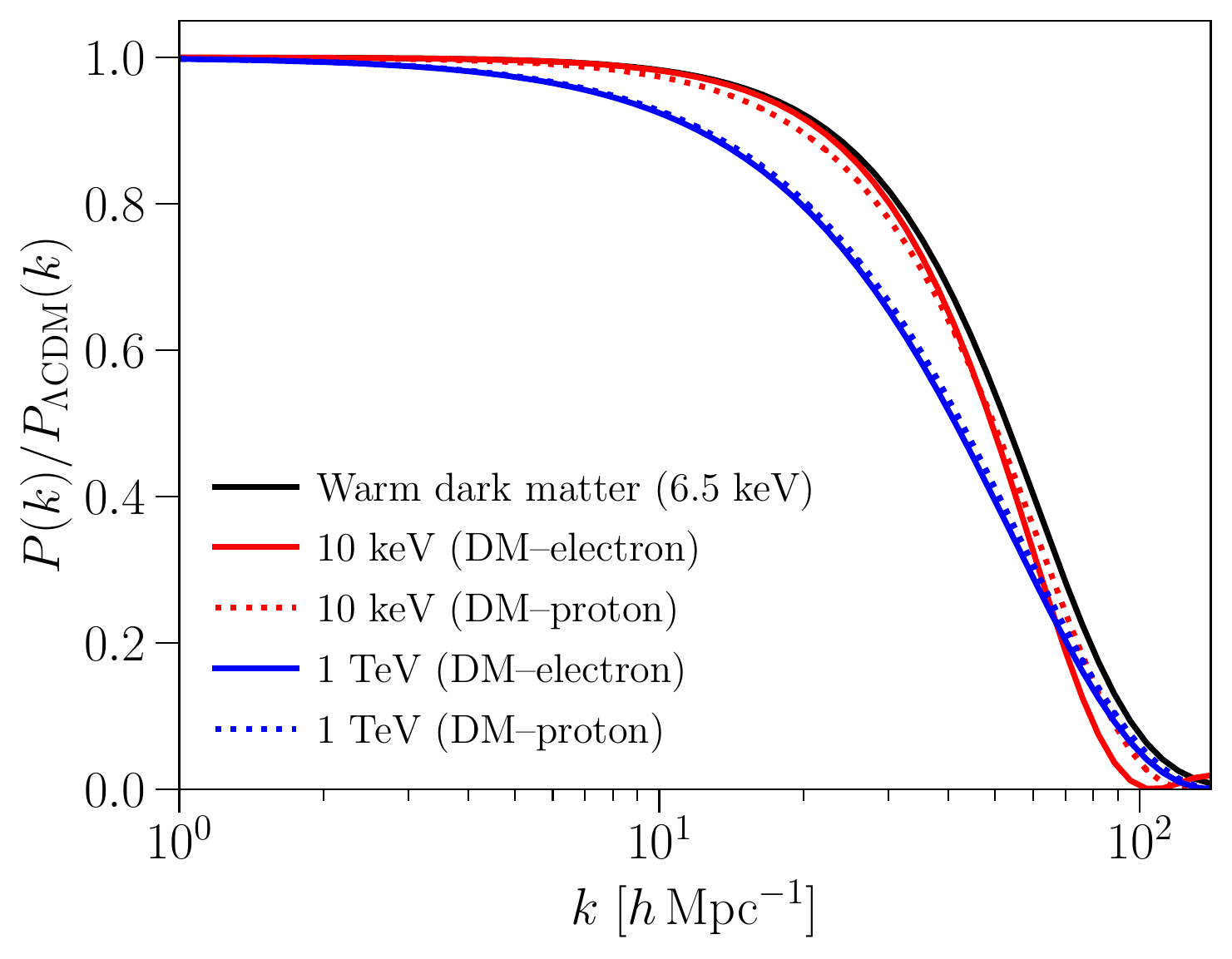}
  \caption{Transfer functions for velocity-independent DM scattering with electrons (solid) and protons (dotted) ruled out at $95\%$ confidence based on the $6.5$ keV thermal relic WDM constraint from Milky Way satellite galaxies (black)~\cite{Nadler:2020prv}.
    We show the transfer functions at two extreme DM masses: $10~\mathrm{keV}$ (red) and $1~\mathrm{TeV}$ (blue).}
  \label{fig:pk}
\end{figure}

In particular, we adopt the same fixed set of cosmological parameters as Ref.~\cite{Maamari:2020aqz} and a maximum wave number of $k=130~h~\mathrm{Mpc}^{-1}$ up to which we ensure that the transfer function suppression is more severe than that of the ruled-out WDM model.
Note that even for the case of $n=0$, we adopt the procedure of Ref.~\cite{Maamari:2020aqz} rather than Ref.~\cite{Nadler:2019zrb} for consistency in our analysis.
Since the transfer function for $n=0$ closely resembles that of WDM, using the procedure in Ref.~\cite{Nadler:2019zrb} yields very similar results.
In Fig.~\ref{fig:pk}, we show the $n=0$ transfer functions for DM scattering with electrons and protons at the ruled-out level for DM masses well above and well below the masses of the electron and proton.

For a large DM mass, the DM--electron scattering and DM--proton scattering transfer functions match closely and exhibit a shallower cutoff than WDM, but our procedure forces the high-$k$ region of the transfer function to follow WDM.
For a low DM mass, the transfer functions closely match the WDM shape near the cutoff region, but the transfer function for DM--electron scattering exhibits a sharper drop at high $k$ compared to the case of DM--proton scattering.
This slight difference in the high-$k$ region between DM--proton and DM--electron scattering produces a different behavior in the mass dependence of the $n=0$ limit at low DM masses.
We expect studies using cosmological simulations with interacting DM would only improve upon these conservative results.
The results of our analysis are summarized in Table~\ref{tab:bounds-mw}, and the limits are plotted in the right panel of Fig.~\ref{fig:limits_all}.

Finally, we do not consider $n<0$ in this part of the analysis, because these models produce scattering in post-recombination universe, and the resulting transfer function cannot be related to that of WDM using the conservative numerical procedure to produce meaningful bounds; the transfer function for these models is illustrated in Ref.~\cite{Boddy:2018wzy}.
A dedicated analysis is required to obtain bounds from satellite abundance measurements for these interaction models, which we leave to future work.

\section{Comparison to direct detection}
\label{sec:compare}

To demonstrate the complementarity between cosmological and low-energy laboratory searches for DM interactions, we compare the upper limits on DM--electron scattering obtained in this work to constraints from electronic recoil direct detection experiments.
Direct detection limits are cast in terms of the quantity~\cite{Essig:2011nj,Essig:2015cda}
\begin{equation}
  \bar{\sigma}_e \equiv \frac{\mu_{\chi e}^2}{16\pi m_\chi^2 m_e^2} \overline{|\mathcal{M} (\alpha m_e)|^2} \, ,
  \label{eq:sigma_e}
\end{equation}
where $\mu_{\chi e} \equiv m_\chi m_e / (m_\chi + m_e)$, $\alpha$ is the fine structure constant, and $\mathcal{M}(\alpha m_e)$ is the matrix element for DM--electron elastic scattering, evaluated at a momentum transfer of $q \equiv |\vec{q}| = \alpha m_e$.
The full matrix element squared for scattering is
\begin{equation}
  \overline{|\mathcal{M} (q)|^2} =
  \overline{|\mathcal{M} (\alpha m_e)|^2} \times
  |F_\textrm{DM}(q)|^2 \ ,
\end{equation}
where the DM form factor $F_\textrm{DM}(q)$ encapsulates the dependence on momentum transfer and the overbar denotes averaging over initial and summing over final spin states.
In the center-of-mass frame, the square of the momentum transfer in the nonrelativistic limit is $q^2 = 2\mu_{\chi e}^2 v^2 (1-\cos\theta)$, where $\theta$ is the scattering angle, and the differential cross section is
\begin{equation}
  \frac{d\sigma}{d\Omega} = \frac{\bar{\sigma}_e}{4\pi} |F_\textrm{DM}(q)|^2 \, .
  \label{eq:relate_sigma_e}
\end{equation}
For various choices of $|F_\textrm{DM}|$, we can relate $\bar{\sigma}_e$ to the coefficient of the momentum-transfer cross section $\sigma_0$, for a given velocity power-law index $n$.

Note that direct detection experiments search for evidence of an ionization signal produced in an inelastic scattering process between DM and an electron bound within an atom; thus, the calculation of the detection rate must also incorporate a form factor for the ionization probability.
Our cosmological analyses in Sec.~\ref{sec:analysis} constrain DM scattering during the pre-recombination era when the Universe is fully ionized, so we are concerned with elastic scattering processes only and Eq.~\eqref{eq:relate_sigma_e} is the appropriate quantity for comparison purposes.
Furthermore, at the time of recombination, there is insufficient kinetic energy to ionize electrons that become bound in atomic hydrogen through direct scattering with DM.

Let us consider a DM form factor parameterized as
\begin{equation}
  |F_\textrm{DM}(q)|^2 = \left(\frac{q}{\alpha m_e}\right)^{n} \, ,
\end{equation}
where $n$ is an integer.
Integrating Eq.~\eqref{eq:relate_sigma_e} according to Eq.~\eqref{eq:sigmaMT}, we find
\begin{equation}
  \sigma_\textrm{MT} = \bar{\sigma}_e \frac{4}{4+n} \left(\frac{2\mu_{\chi e}}{\alpha m_e}\right)^n v^n \,
\end{equation}
for $n > -4$.
We identify the prefactor of $v^n$ in the above equation with $\sigma_0$ in Eq.~\eqref{eq:sigmaMT-param}; therefore, we can immediately relate our results to those from direct detection through a simple rescaling for various values of $n$.

The three cases often considered in the direct detection literature are $F_\textrm{DM} \in \{1, \alpha m_e / q, (\alpha m_e / q)^2 \}$ (see e.g., Ref.~\cite{Essig:2015cda}), which relate to our results on $\sigma_0$ for $n \in \{0,-2,-4\}$, respectively.
For the case of $n=-4$, the integral to calculate $\sigma_\textrm{MT}$ has a logarithmic divergence in the limit of far-forward scattering (as $\theta \to 0$), and we may regulate the divergence with a small-angle cutoff $\theta_D \ll 1$.
Thus, we find the correspondence between $\sigma_0$ and $\bar{\sigma}_e$ is
\begin{equation}
  \sigma_0 = \bar{\sigma}_e \times
  \begin{cases}
    1 & \textrm{for } n=0 \\
    \frac{\alpha^2 m_e^2}{2\mu_{\chi e}^2} & \textrm{for } n=-2 \\
    \left(\frac{\alpha^2 m_e^2}{2\mu_{\chi e}^2} \right)^2 \ln\left(\frac{2}{\theta_D}\right) & \textrm{for } n=-4 \, .
  \end{cases}
\end{equation}
We interpret the cutoff angle for $n=-4$ in the context of millicharged DM: due to Debye screening of electromagnetic fields in a plasma, the cutoff angle is $\theta_D = m_D / (\mu_{\chi e}v)$, where the Debye mass is $m_D = \sqrt{4\pi\alpha n_e / T_\gamma}$.

The Debye logarithm $\ln(2/\theta_D)$ introduces two complications: the momentum-transfer cross section has a $v$ dependence that is not captured by a power-law scaling, and $m_D  \approx 8.1 \times 10^{-16} (1+z)~\mathrm{MeV}$ has a redshift dependence that renders $\sigma_0$ an evolving quantity rather than a constant.
However, $\sigma_\textrm{MT}$ has only a logarithmic dependence on $\theta_D$, and the Debye logarithm varies at the level of tens of percent over the redshift range of interest for \textit{Planck}.
Moreover, \textit{Planck} data seem to provide the greatest constraining power on $\sigma_0$ near redshift $z_\diamond=2\times 10^4$: the data have a high signal-to-noise for multipoles $\ell \sim 1400$, roughly corresponding to perturbation modes that enter the sound horizon around this time~\cite{Boddy:2018kfv}.
We neglect the impact of the evolution of $m_D$ and $v$, fixing the Debye logarithm at the reference redshift $z=z_\diamond$ and fixing $v^2$ to its approximated thermal value, given by the right-hand side of Eq.~\eqref{eq:modification}.
To fix these quantities, we set the $\Lambda$CDM parameters to their best-fit values from the \textrm{Planck} 2018 TTTEEE+lowE+lensing analysis~\cite{Aghanim:2018eyx} and the DM scattering parameter $\sigma_0$ to be its 95\% C.L.\ upper limit in Table~\ref{tab:bounds-cmb} for each corresponding DM mass.

\begin{figure}[!t]
  \includegraphics[width=0.95\linewidth]{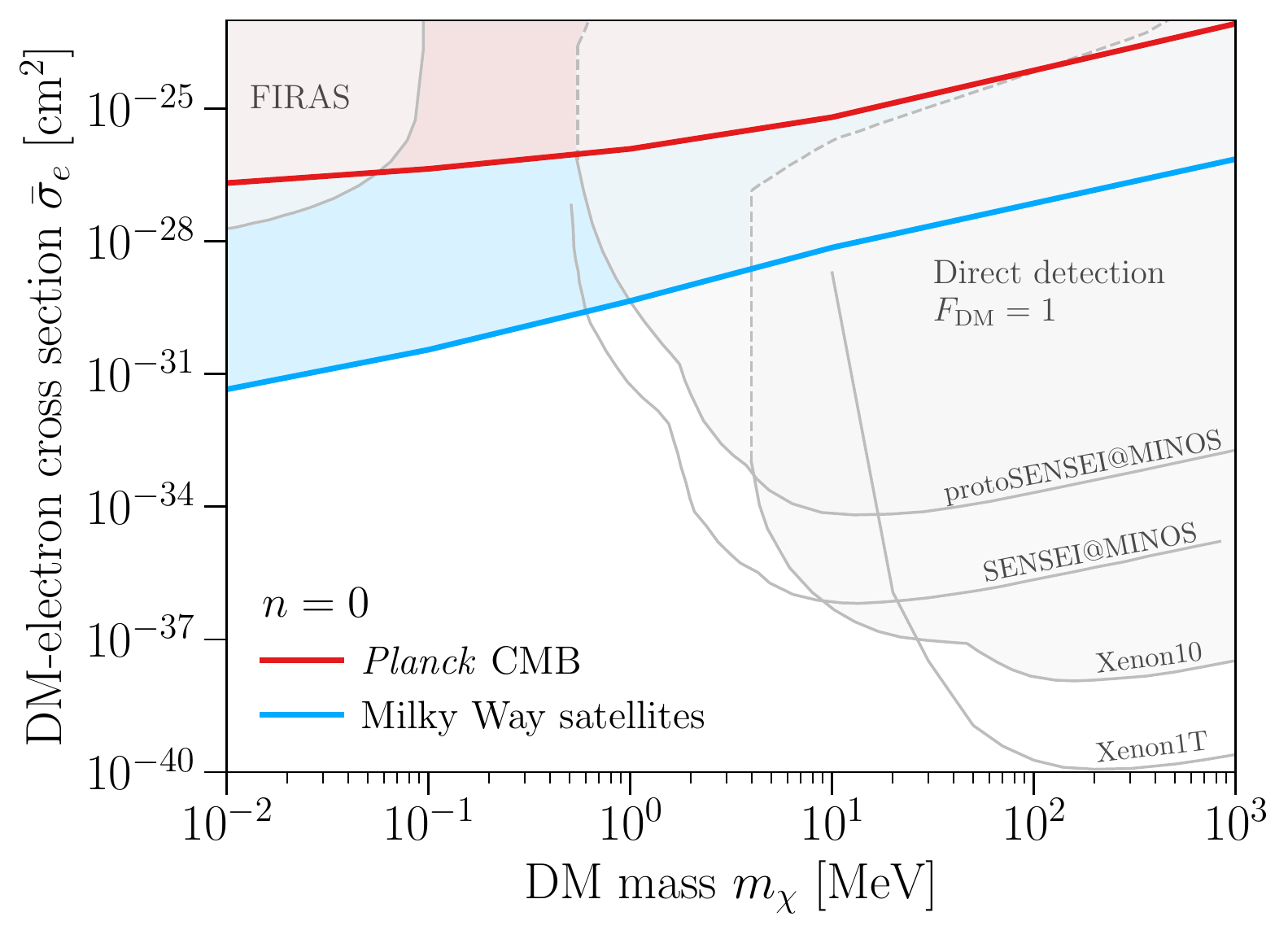} \\
  \includegraphics[width=0.95\linewidth]{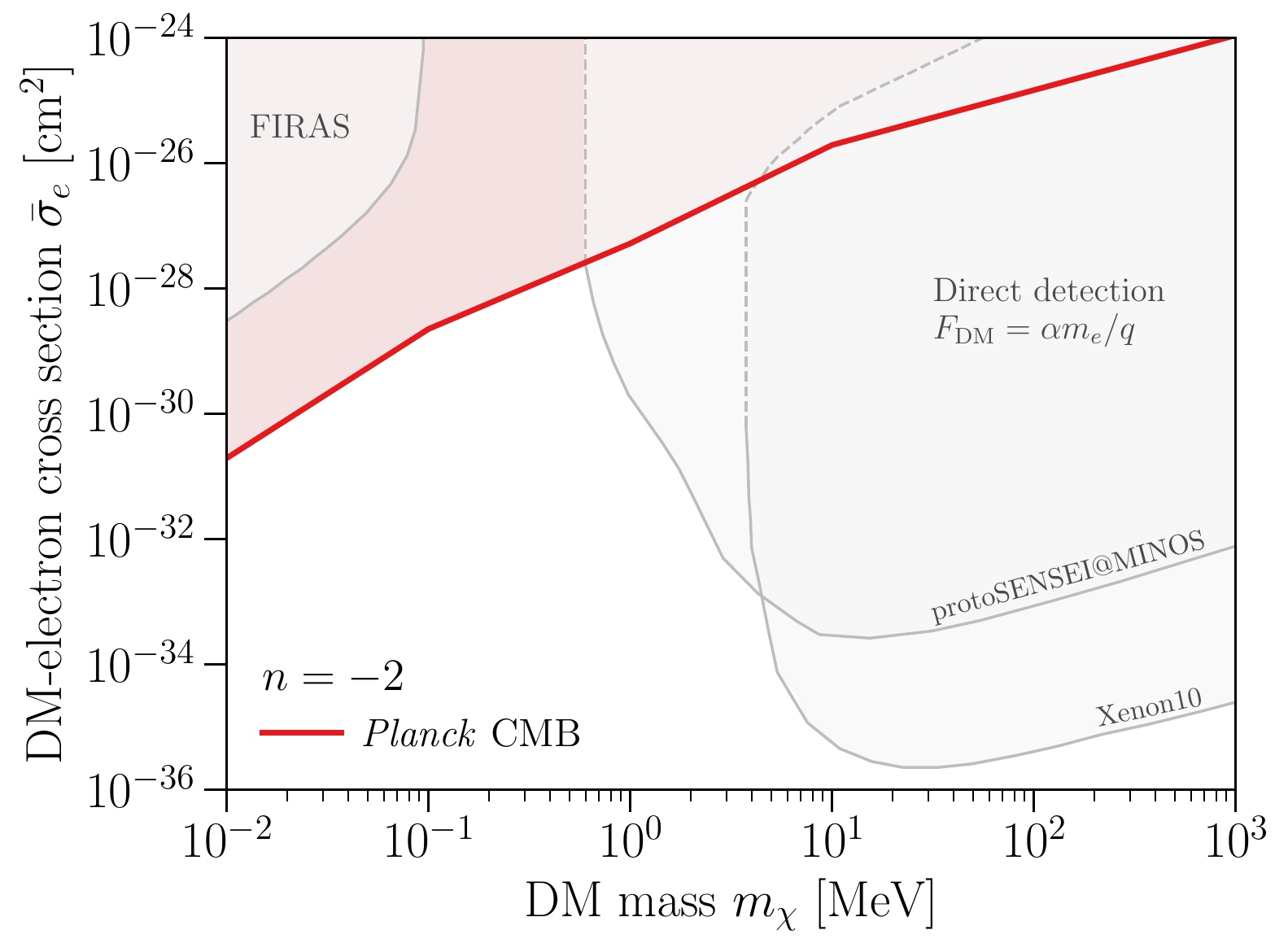} \\
  \includegraphics[width=0.95\linewidth]{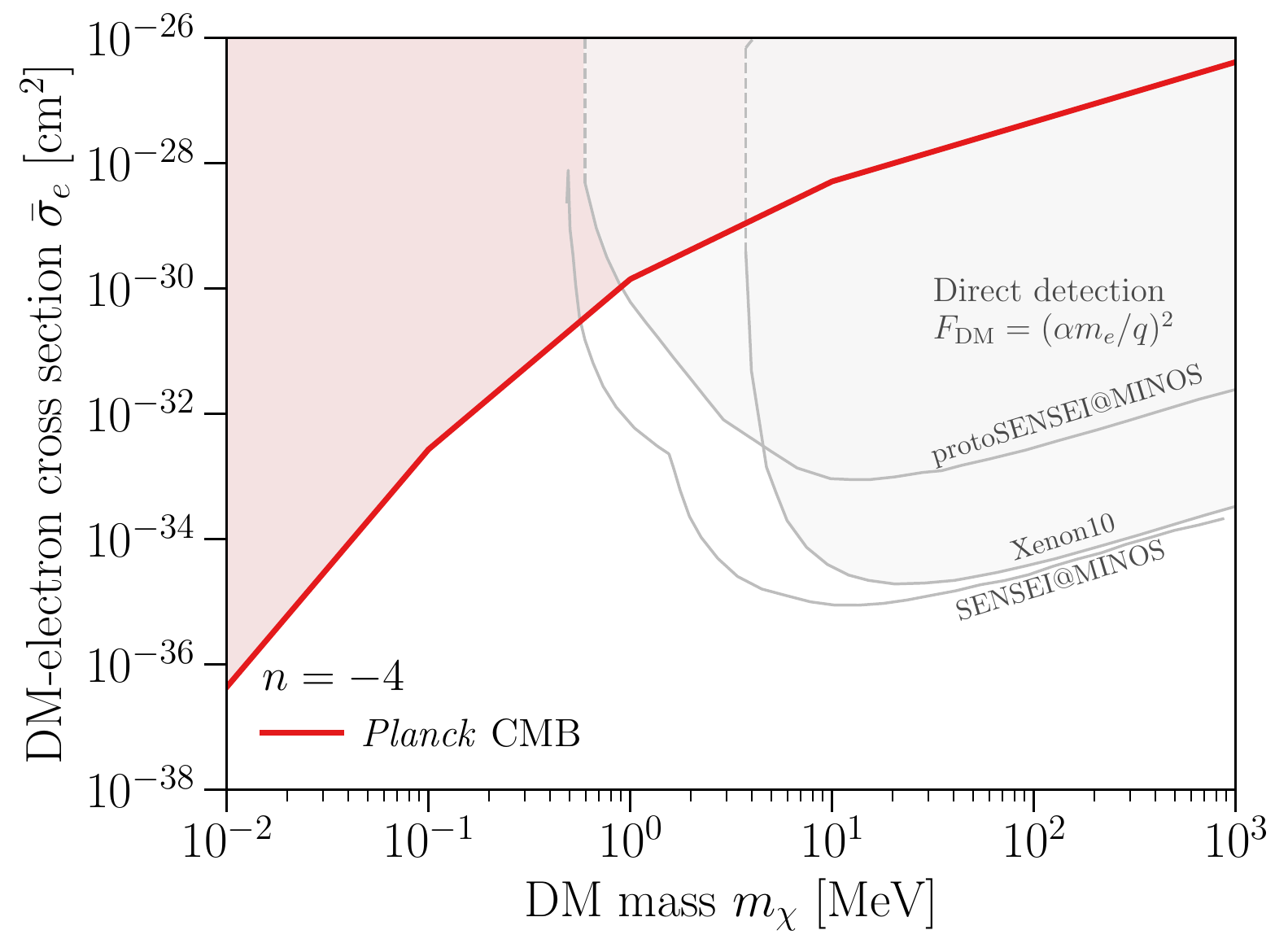}
  \caption{Comparison of the CMB (red) and Milky Way satellite (blue) results from this work and the exclusion bounds from electronic-recoil direct detection experiments and FIRAS spectral distortions (gray)~\cite{Ali-Haimoud:2021lka}.
    We show recent direct detection bounds from Xenon1T~\cite{Aprile:2019xxb} and SENSEI@MINOS~\cite{Barak:2020fql} for $F_\textrm{DM}=1$ (top), $F_\textrm{DM}=\alpha m_e / q$ (middle), and $F_\textrm{DM}=(\alpha m_e / q)^2$ (bottom); shaded regions for Xenon10~\cite{Essig:2017kqs} and protoSENSEI@MINOS~\cite{Abramoff:2019dfb} incorporate ceiling calculations from Ref.~\cite{Emken:2019tni}.
    We translate available cosmological limits for $n=0$ (top), $n=-2$ (middle), and $n=-4$ (bottom) to the quantity $\bar{\sigma}_e$ defined in Eq.~\eqref{eq:sigma_e}.}
  \label{fig:limits_dd}
\end{figure}

In Fig.~\ref{fig:limits_dd}, we compare our observational bounds with exclusion bounds from electronic-recoil direct detection experiments in the parameter space of $\bar{\sigma}_e$ versus DM mass $m_\chi$.
The top, middle, and bottom panels of the figure correspond to the $n=0$ [$F_\textrm{DM}=1$], $n=-2$ [$F_\textrm{DM}=\alpha m_e / q$], and $n=-4$ [$F_\textrm{DM}=(\alpha m_e / q)^2$] cases, respectively.
The shaded gray regions show bounds from the Xenon10~\cite{Essig:2017kqs} and protoSENSEI@MINOS~\cite{Abramoff:2019dfb} direct detection experiments, as presented in Ref.~\cite{Emken:2019tni}, which includes calculations of the sensitivity ceilings (shown as dashed gray lines).
Additionally, we show more recent direct detection bounds from Xenon1T~\cite{Aprile:2019xxb} for $n=0$ and from SENSEI@MINOS~\cite{Barak:2020fql} for $n=0$ and $n=-4$ as individual gray lines with no shading (Ref.~\cite{Emken:2019tni} includes ceiling projections for SENSEI, but we do not include them here).
We include limits (also shaded gray regions) from $\mu$-type spectral distortions using FIRAS data for $n=0$ and $n=-2$ from Ref.~\cite{Ali-Haimoud:2021lka}.

The results of this work exclude new regions of DM parameter space, particularly for cross sections above direct detection sensitivity ceilings and for DM masses below direct detection mass sensitivity thresholds.
For $n=0$, our CMB constraint bridges the gap between limits at low DM masses from spectral distortions and limits at high DM masses from direction detection, while our Milky Way satellite constraint is stronger than both spectral distortions and CMB.
For $n=-2$, our CMB constraint is stronger than spectral distortion bounds.
Finally, for $n=-4$, our CMB analysis provides the only cosmological constraint on DM--electron scattering and excludes a large region of parameter space at small DM masses.

\section{Summary and Discussion}
\label{sec:conclusions}

This work presents new observational constraints on elastic DM--electron scattering using its effects on the matter distribution in the Universe. Specifically, we rely on the latest CMB measurements from \textit{Planck} and measurements of the abundance of Milky Way satellite galaxies detected by the DES and Pan-STARRS1 surveys.
To explore the space of possible DM scattering models, we parameterize the momentum-transfer cross section as a power law of the relative particle velocity, with power-law indices $n\in \{-4,-2,0,2,4,6\}$.
Interaction models with a negative power-law index lead to momentum exchange between DM and baryons primarily at late times, while for non-negative values of $n$, the primary effects of scattering take place in the early Universe.
We constrain all of these scenarios using the CMB data; when using the satellite abundance, the existing analysis methods rely on relating the shape of the transfer function to that of WDM, so we limit our analyses to non-negative values of $n$ that feature early-time scattering only, where these methods are applicable.

Our resulting bounds are presented in Fig.~\ref{fig:limits_all}.
For the case of $n\geq 0$, where well-defined comparisons exist, Milky Way satellite abundances are more constraining than the CMB, because they probe matter clustering on smaller scales that are more strongly affected by DM--electron interactions in the early Universe.
Our constraints for $n<0$ from CMB data present some of the strongest observational bounds to date; we defer a systematic exploration of the effects of these models on nonlinear structure, including Milky Way satellite abundances, for future work.

We note that recent joint analyses of small-scale structure probes have achieved more stringent WDM constraints, for example by combining strong gravitational lensing and Milky Way satellites~\cite{Nadler:2021dft,Enzi:2020ieg}.
Different small-scale structure tracers are sensitive to both the abundances and concentrations of low-mass halos in distinct ways, precluding a straightforward mapping to interacting DM models that may impact the corresponding observables differently.
A joint analysis of the effects of DM interactions on multiple small-scale structure probes is an interesting direction for future work.

We also compare our bounds with the constraints from electronic-recoil direct detection experiments in Fig.~\ref{fig:limits_dd}.
We find that observational bounds, especially those that involve small-scale tracers like satellite galaxies, have overlap with the upper limits obtained from direct detection.
Moreover, our bounds present the strongest observational limits on sub-MeV DM interactions with electrons.
They also conclusively exclude regions of the parameter space above the detection ceiling of direct detection experiments.

There are astrophysical limits on DM--electron scattering that arise from constraints on the cooling of supernovae~\cite{Guha:2018mli}, DM capture in the Sun~\cite{Garani:2017jcj}, and direct detection of low-mass DM that undergoes cosmic ray upscattering~\cite{Cappiello:2018hsu}.
Since these analyses either rely upon a DM annihilation signal or work in the relativistic scattering regime, neither of which pertain to our analyses, we do not include them in Fig.~\ref{fig:limits_dd}.
There may also be bounds on the mass of DM from contributions to the energy density of relativistic species at Big Bang nucleosynthesis (BBN); however, these bounds depend on the spin statistics of the DM particle~\cite{Boehm:2013jpa,Nollett:2014lwa,Krnjaic:2019dzc} and may be circumvented in certain DM scenarios~\cite{Berlin:2017ftj}.

Finally, we expect that the same methods we employ in this study, which have enabled some of the leading observational constraints on DM elastic scattering with electrons and protons, may be applied to other data sets as well.
For example, both CMB experiments such as the Simons Observatory~\cite{SimonsObservatory:2018koc} and surveys like the Rubin Observatory Legacy Survey of Space and Time~\cite{LSST:2008ijt} deliver their first data in the coming years. Combined with our theoretical framework, these data may enable  searches for DM interactions throughout cosmic history.

\begin{acknowledgments}
V.~G.\ is supported by the National Science Foundation under Grant No.~PHY-2013951.
This research received support from the National Science Foundation (NSF) under grant No.\ NSF DGE-1656518 through the NSF Graduate Research Fellowship received by E.~O.~N.
\end{acknowledgments}

\appendix
\section{Scattering with protons}
\label{sec:app}

In this appendix, we present results for a CMB analysis of DM scattering with protons.
Deriving limits on DM--proton scattering using \textit{Planck} data has been considered in previous literature~\cite{Dvorkin:2013cea,Gluscevic:2017ywp,Boddy:2018kfv,Xu:2018efh,Slatyer:2018aqg,Boddy:2018wzy}.
The modified \texttt{CLASS} code used for this work is based on the code used in Refs.~\cite{Gluscevic:2017ywp,Boddy:2018kfv,Boddy:2018wzy}; various improvements have made the code more numerically stable and ready for public release.
With these improvements, we are able to explore certain regions of DM parameter space that had numerical difficulties in our previous studies.
Therefore, we revisit the scenario of DM--proton scattering both to cover these missed regions of parameters space and to serve as a consistency check, aiding in the validation of our code.

\begin{figure}[t]
  \includegraphics[width=0.95\linewidth]{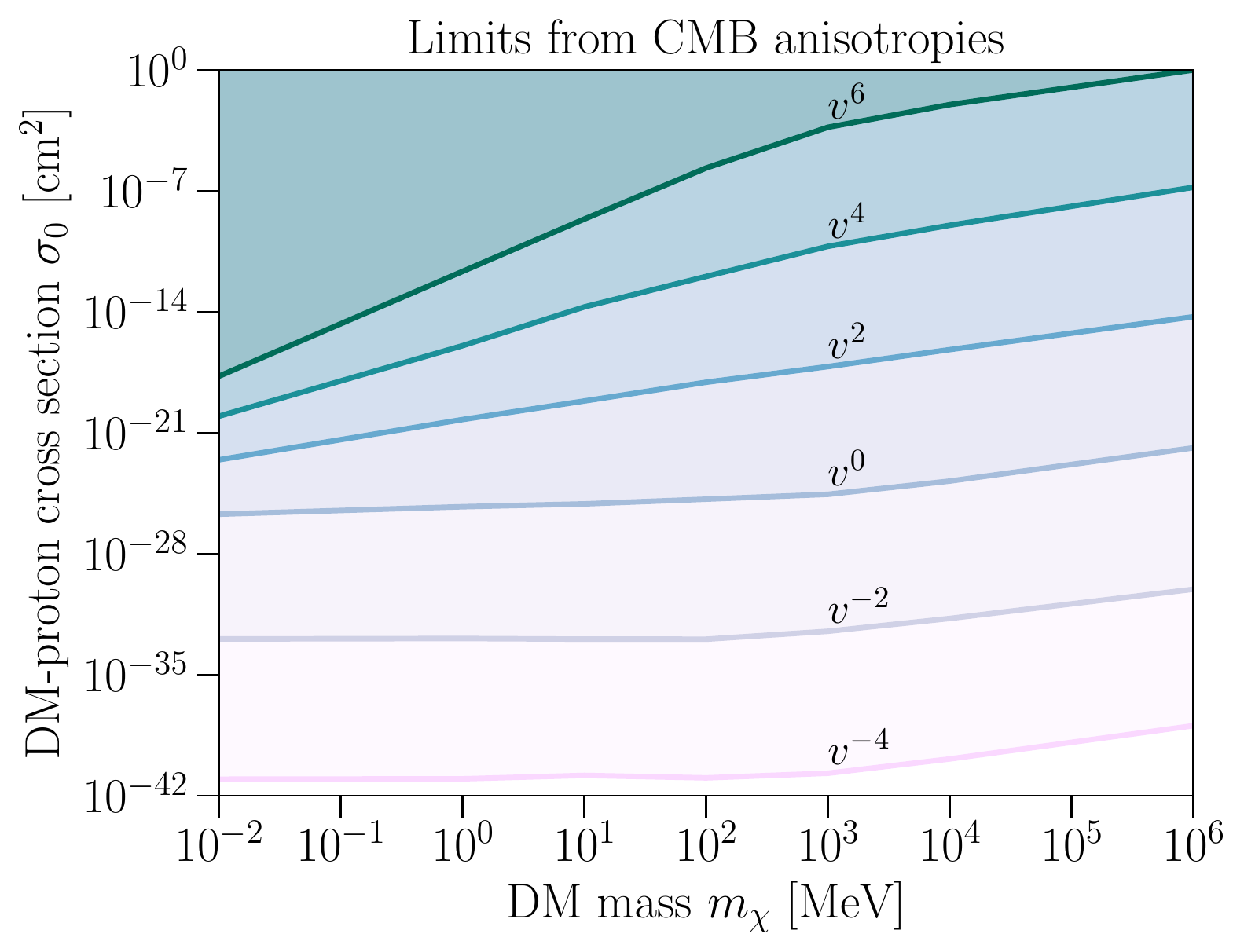}
  \caption{The 95\% C.L.\ upper limits on $\sigma_0$, the coefficient of the momentum-transfer cross section for DM--proton scattering, as a function of DM mass.
    The cross section scales as $v^n$, as indicated in the plots, where $v$ is the relative velocity of scattering particles.
    The shaded region above each line is excluded by \textit{Planck} 2018 CMB temperature, polarization, and lensing power spectra.}
  \label{fig:limits_all_protons}
\end{figure}

We consider scattering with protons in the form of neutral or ionized hydrogen; we neglect scattering with helium.
Our analysis follows the same procedure outlined in Sec.~\ref{sec:analysis-cmb} for DM--electron scattering.
The 95\% C.L.\ upper limits on $\sigma_0$, where $\sigma_0$ now refers to the coefficient of the momentum-transfer cross section for DM--proton scattering, are shown in Fig.~\ref{fig:limits_all_protons} and listed in Table~\ref{tab:bounds-cmb-protons}.

The analysis for DM--proton scattering using Milky Way satellite abundances was performed in Ref.~\cite{Maamari:2020aqz} with the same modified \texttt{CLASS} code used in this work.
Therefore, we refer the reader to Ref.~\cite{Maamari:2020aqz} for these bounds.

\begin{table}[!h]
  \centering
  \begin{tabular}{|r!{}l|c|c|c|c|c|c|}
    \hline
    \multicolumn{2}{|c|}{\textbf{DM}} & \multicolumn{6}{c|}{$n$} \\
    \cline{3-8}
    \multicolumn{2}{|c|}{\textbf{mass}} & $-4$ & $-2$ & $0$ & $2$ & $4$ & $6$ \\
    \hline
    1 & keV & 8.9e-42 & 1.1e-33 & 1.2e-26 & 1.9e-24 & 8.4e-23 & 1.8e-21 \\
    \hline
    1 & MeV & 9.5e-42 & 1.3e-33 & 5.3e-26 & 5.9e-21 & 1.1e-16 & 2.2e-12 \\
    \hline
    10 & MeV & 1.5e-41 & 1.2e-33 & 7.7e-26 & 7.0e-20 & 1.9e-14 & 2.4e-09 \\
    \hline
    100 & MeV & 1.1e-41 & 1.1e-33 & 1.5e-25 & 8.5e-19 & 1.1e-12 & 2.1e-06 \\
    \hline
    1 & GeV & 2.0e-41 & 3.3e-33 & 2.8e-25 & 6.8e-18 & 6.2e-11 & 4.9e-04 \\
    \hline
    10 & GeV & 1.4e-40 & 1.8e-32 & 1.6e-24 & 6.6e-17 & 1.0e-09 & 9.9e-03 \\
    \hline
    1 & TeV & 1.1e-38 & 8.9e-31 & 1.4e-22 & 5.2e-15 & 1.6e-07 & 9.9e-01 \\
    \hline
  \end{tabular}
  \caption{The 95\% C.L.\ upper limits on $\sigma_0$, the coefficient of the momentum-transfer cross section for DM--proton scattering, in units of $\mathrm{cm}^2$ from the CMB analysis of Sec.~\ref{sec:app} and shown in Fig.~\ref{fig:limits_all_protons}.}
  \label{tab:bounds-cmb-protons}
\end{table}

\bibliography{dme}

\end{document}